\documentclass[prl,aps,reprint]{revtex4-1}
\usepackage{multirow}
\usepackage[table]{xcolor}
\usepackage{rotating}
\usepackage{amsmath}
\usepackage{bm}
\usepackage{array}
\usepackage{graphicx}
\usepackage{dcolumn}   
\usepackage{amssymb}
\usepackage[english]{babel}

\newcommand{\Th}{\mathbf{\hat{t}}}

\newcommand{\erho}{\mathbf{\hat{e}}_{\rho}}

\newcommand{\rc}{\mathbf{r}}
\newcommand{\R}{\mathbf{R}}
\newcommand{\tR}{\mathbf{\tilde{R}}}

\newcommand{\Su}{\mathbf{S}}
\newcommand{\f}{\boldsymbol{\mathfrak{f}}}

\begin{document}
\title{Tubular-body theory for viscous flows}
\date{\today}
\author{Lyndon Koens\footnote{lyndon.koens@mq.edu.au, L.M.Koens@hull.ac.uk}}
\affiliation{ Department of Mathematics \& Statistics, Macquarie University, Sydney, NSW 2113, Australia}
\affiliation{ Department of Physics \& Mathematics, University of Hull, Hull HU6 7RX, UK}

\begin{abstract}
Cable-like bodies play a key role in many interdisciplinary systems but are hard to simulate. Asymptotic theories, called slender-body theories, are effective but apply in specific regimes and can be hard to extend beyond leading order. In this letter we develop an exact slender-body-like theory for the surface traction of cable-like bodies in viscous flow. This theory expresses the traction as a series of solutions to a well-behaved one-dimensional Fredholm integral equation of the second kind. This process can be simply generalised to other systems.

\end{abstract}
\maketitle
\def\v{\vspace{2cm}}

\textit{Introduction}- Wiry objects are important to many systems: Spermatozoa and bacteria actuate slender appendages called flagella to swim \cite{Lauga2016, Gaffney2011}; Eukaryotic cells change and maintain their shape with microtubules and actin filaments \cite{Nazockdast2017a}; Clays are colloids of electrically-charged ribbons \cite{Ruzicka2011}; and fibre-reinforced plastics are light weight meta-materials that can be used for machine parts \cite{Sanjay2018}. Systems with cable-like bodies often display complex and emergent behaviours but have internal structures that are hard to probe experimentally \cite{Lemma2021}. As a result theoretical and numerical models are needed to complement the experiments and improve understanding.

Unfortunately, the direct numerical simulation of tube-like bodies is often computationally heavy \cite{Koens2021a}. This is because the thickness of each body may be much smaller than its length and so a high resolution is required. Asymptotic theories, called slender-body thoeries (SBT), have been developed to overcome this. These models exploit the separation in length scales to create an approximation for the behaviour \cite{Kim2005}. This process often reduces the model to a system of one dimensional equations, thereby also increasing the computational efficiency. 

SBT has been used in many fields with one of the most successful examples coming from slow-viscous flows. SBT for viscous flows has accurately modelled microscopic swimmers \cite{Lauga2016, Gaffney2011}, flexible filaments \cite{DuRoure2019} and settling rods \cite{Tornberg2004}. In its most popular form, it expresses the force per unit length on an isolated filament through one dimensional integral equation \cite{Keller1976a, Johnson1979, Koens2018, Gotz2000}. Though the kernel of this operator is singular, this singularity asymptotically cancels with a `local' term. Even so, this model can be difficult to implement numerically and is known to suffer from high eigenvalue instabilities \cite{Gotz2000, Tornberg2004, Andersson2020}.

The approximate nature of SBTs typically restricts their use to specific regimes. Most assume that the thickness of the body is much less than all other length scales in the problem \cite{Barta1988, Tornberg2004, Gueron1992, Higdon2006}, though some restrict certain lengths to be similar to or smaller than the thickness \cite{Barta1988, Man2016, Brennen1977a}. Only one example exists without such a condition to the authors knowledge \cite{Koens2021a}. However it restricts the geometry to that of a rod perpendicular to a wall. Furthermore most SBTs are hard to extend beyond leading order \cite{Johnson1979, Katsamba2020}. Yet effective models without such limitations are needed. Realistic systems are dynamic and complex, with wiry bodies changing shape, interacting with walls, and each-other. This has been identified as a key issue in interdisciplinary research within reviews \cite{DuRoure2019, Brennen1977a, Kugler2020} and a Nature Physics comment \cite{Reis2018}.

In this letter we develop tubular-body theory (TBT) for viscous flow. TBT is a SBT-like theory which exactly determines the traction on the surface of cable-like bodies. The theory is found by adding and subtracting the solution to an effective sphereoid to the singular-layer boundary integral representation and then expanding the system using a binomial series. Rearranging these equations, the traction is expressed as a series of solutions to a one-dimensional Fredholm integral equation of the second kind with a compact and self-adjoint kernel. This enables TBT to retain many of the computational advantages of SBT without the limitations. These equations are found to capture the behaviour of spheroids and tori well outside classical SBT limits and simulate the motion of a tightly wound helix.

\begin{figure}
\centering
\includegraphics[width=0.45\textwidth]{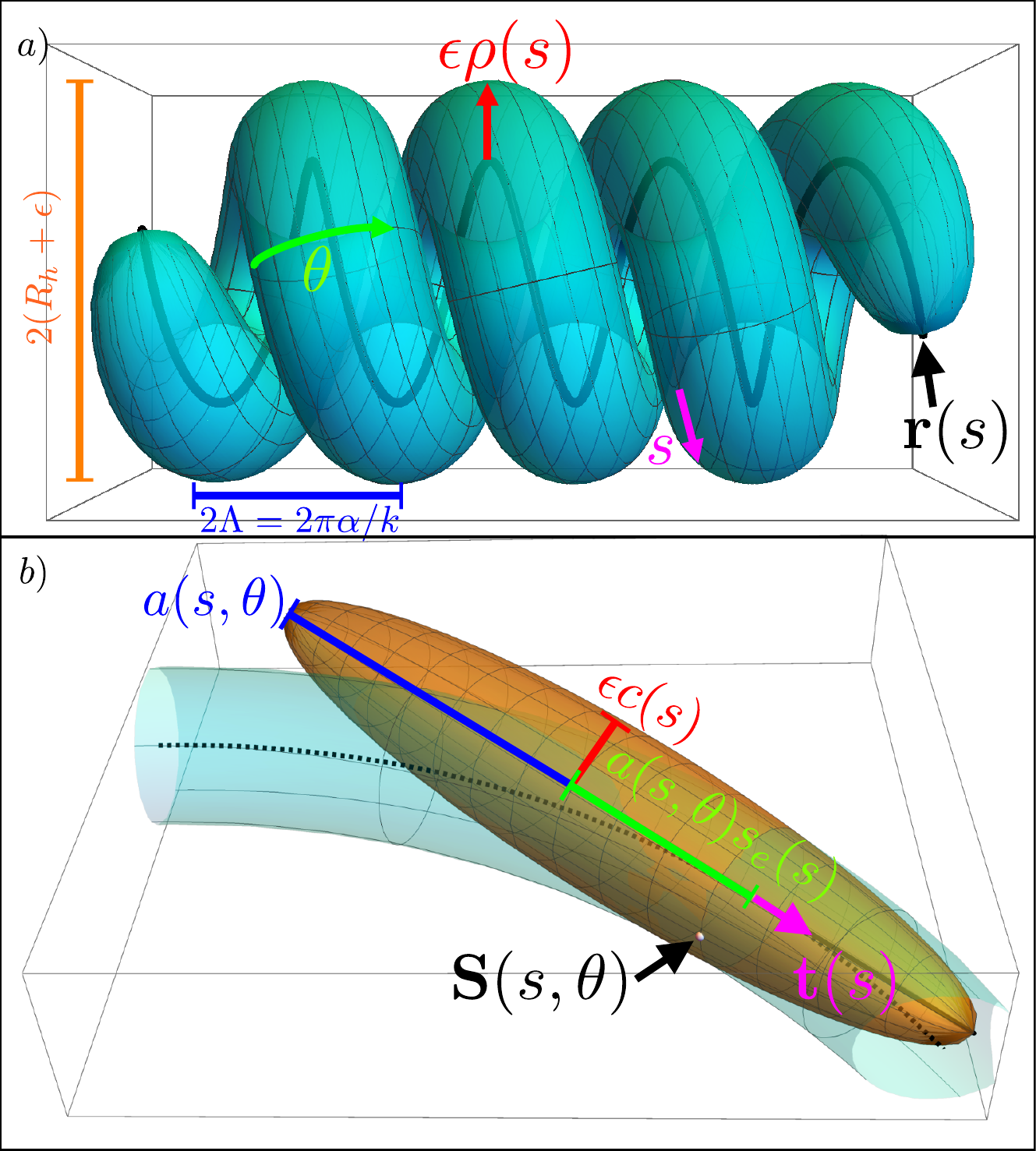}
\caption{a) A tightly wound helix has a locally tubular structure. $\rc(s)$ (black) is the centreline, $\epsilon \rho(s)$ (red) is the thickness, $\Lambda$ (blue) is the helix pitch, $ R_h $ (orange) is the helix radius. Plot uses $\epsilon=0.05$, $\Lambda = 1.1 \epsilon$ and $R_h=1.5\epsilon$. { b) The effective spheroid used to replicate the surface around $\Su(s,\theta)$. $a(s,\theta)$ (blue) is the distance from the centre to pole along the symmetry axis, $\epsilon c(s)$ (red) is the equatorial radius, $a(s,\theta) s_e(s)$ (green) is the distance from the centre to $\mathbf{r}(s)$ along the tangent, $\Th(s)$, (magenta). These parameters are chosen such that the surface point and tangent plane matches at $\Su(s,\theta)$.}} 
\label{fig:ex}
\end{figure}

\textit{Tubular-body theory}- Consider the behaviour of tubular bodies in incompressible viscous flow (Fig.~\ref{fig:ex}a). The surface of these bodies can be written as
\begin{equation}
\Su(s,\theta) = \rc(s) + \epsilon \rho(s) \erho(s,\theta),
\end{equation}
where $\rc(s)$ is the centreline of the body parametrised by the arclength $s \in[-1,1]$, $\epsilon \rho(s) \in [0,1]$ is the thickness of the body at $s$, $\erho(s,\theta)$ is the local radial vector perpendicular to the centreline tangent $\partial_s \rc(s) = \Th(s)$. Without any loss of generality we choose $\erho(s,\theta)$ to be parametrised such that $\partial_s \erho(s,\theta) = - \kappa(s) \cos(\theta- \int_0^{s} \tau(s) \,ds) \Th(s)$ \cite{Koens2018}, where $\kappa(s)$ and $\tau(s)$ are the curvature and torsion of the centreline, respectively. {This parametrisation does not allow for jumps in $\rho(s)$ or blunt ends like in many SBTs. }

Assuming that the body does not change its volume, the flow around the body, $\mathbf{u}(\mathbf{x})$, can be described through the single-layer boundary integral representation for incompressible Stokes flow. For an isolated body this representation is given by
\begin{equation}
8 \pi \mu   \mathbf{u}(\mathbf{x})= \int_{-1}^{1} \,ds' \int_{-\pi}^{\pi} \,d\theta' \left(\frac{\mathbf{I}}{|\R'|} + \frac{\R' \R'}{|\R'|^3} \right) \cdot \f(s',\theta'), \label{SLBIf}
\end{equation}
where $\R' = \mathbf{x} - \Su(s',\theta')$ is a vector from the surface to the point of interest in the fluid, $\mathbf{I}$ is the identity matrix, $\mu$ is the viscosity of the fluid and $\f(s,\theta)$ is the fluid traction on the body multiplied by the integration surface element. This equation determines the flow both inside and outside the body. In the limit $\mathbf{x} \to \Su(s,\theta)$, Eq.~\eqref{SLBIf} becomes 
\begin{equation}
8 \pi \mu  \mathbf{U}(s,\theta)= \int_{-1}^{1} \,ds' \int_{-\pi}^{\pi} \,d\theta' \left(\frac{\mathbf{I}}{|\R|} + \frac{\R \R}{|\R|^3} \right) \cdot \f(s',\theta'), \label{SLBI}
\end{equation}
where $ \mathbf{U}(s,\theta)$ is the surface velocity of the body, and $\R = \Su(s,\theta) - \Su(s',\theta')$. The above equation is
a well-studied integral equation for $\f(s',\theta')$ \cite{Pozrikidis1992}.

The integrand of the above equation blows up when $(s',\theta') =(s,\theta)$. This divergence can be removed by adding and subtracting another geometry with a known solution that cancels with the kernel at said point. This process is reminiscent to that used by Batchelor in the development of his SBT \cite{Batchelor2006}. We chose to add and subtract the flow around a translating spheroid with { symmetry axis aligned with the body tangent, $\Th(s)$, at $s$. This spheroid has three unset geometric parameters: the distance from the centre to pole along the symmetry axis, $a$,  the equatorial radius, $\epsilon c$, and the distance from the centre to $\mathbf{r}(s)$, $ a s_e $, (Fig.~\ref{fig:ex}b). These parameters should be chosen to match the tubular-body surface and tangent plane at $(s',\theta') =(s,\theta)$ to provide the best regularisation of the integral kernel. This gives the effective spheroid}
 the parametrisation
\begin{equation}
{ \Su_e(s',\theta',s,\theta)}= {\rc_{e}(s',s, \theta)}  + \epsilon {\rho_e(s',s)} \erho(s,\theta') + \rc(s)  \label{effective}
\end{equation}
where $\rc_e(s',s,\theta) =a(s,\theta) \left[s' - s_e(s) \right]\Th(s)$, $\rho_e(s',s) = c(s) \sqrt{1-s'^2}$, $2 c(s)^2 = \rho^2(s) + \rho(s) \sqrt{\rho^2(s) +4 (\partial_s\rho(s))^2}$, $a(s,\theta) = 1 - \Th(s) \cdot\partial_s \erho(s,\theta)$ and $ s_e(s) =  \rho(s) \partial_s\rho(s) /c^2$. This geometry corresponds to a prolate spheroid if $\alpha =\epsilon c/a <1$  and an oblate spheroid if $\alpha=\epsilon c/a>1$.  It satisfies $\Su_e(s_e,\theta) =\Su(s,\theta)$, $\partial_{s_e}\Su_e(s_e,\theta) =\partial_s\Su(s,\theta)$ and  $\partial_\theta\Su_e(s_e,\theta) =\partial_\theta\Su(s,\theta)$, thereby replicating the position and tangent plane to the surface at $(s,\theta)$. Furthermore it relaxes the typical SBT assumption on the curvature ($ \epsilon \kappa \ll 1$) and assumes that the cross-section satisfies $\rho(s) \partial_s \rho(s) \to$ constant as $\partial_s \rho(s) \to \infty$. {The latter condition allows TBT to model rapidly changing cross-sections provided $\partial_s \rho(s) \neq \infty$ away from ends and different types of ends that satisfy the criteria.} The traction, multiplied by the surface element, from the translational motion of a spheroid in a viscous fluid is constant for the above parametrisation \cite{Brenner1963, Martin2019} and the total force from the motion is known. Hence the boundary integral equation, Eq.~\eqref{SLBI}, after adding and subtracting the effective geometry, can be written as
\begin{eqnarray}
8 \pi \mu \mathbf{U}(s,\theta)&=& \int_{-1}^{1} \,ds' \int_{-\pi}^{\pi} \,d\theta' \left[ \mathbf{G} \cdot \f(s',\theta') -\mathbf{G}_{e} \cdot \f(s,\theta) \right] \notag \\
&& + {\mathbf{M}_{A}'(s,\theta)}\cdot \f(s,\theta),  \label{aSLBI}
\end{eqnarray}
where  $\R_e = {\Su_e(s_e(s),\theta,s,\theta)} - {\Su_e(s',\theta',s,\theta)}$, $\beta{(s,\theta)} =\alpha^2{(s,\theta)}-1$,  
\begin{eqnarray}
\mathbf{G}_{(e)}{(s',\theta',s,\theta)} &=&\frac{\mathbf{I}}{|\tR_{(e)}|} + \frac{\R_{(e)} \R_{(e)}}{|\tR_{(e)}|^3}, \\
{ \mathbf{M}_{A}'(s,\theta) } &=& {\zeta_{\parallel}'(s,\theta) \Th(s) \Th(s) +{\zeta_{\perp}'(s,\theta)} [\mathbf{I}-\Th(s)\Th(s)]} \notag \\ \\
\frac{a \beta^{3/2}}{4 \pi}\zeta_{\parallel}'{(s,\theta)} &=& \left[ (\beta-1) \arccos\left(\alpha^{-1}\right) + \sqrt{\beta}\right],  \\
\frac{a \beta^{3/2}}{2 \pi}\zeta_{\perp}'{(s,\theta)} &=&   \left[(3 \beta +1) \arccos\left(\alpha^{-1}\right) -\sqrt{\beta}\right],
\end{eqnarray}
and the subscript $(e)$ means the notation applies for both the tubular body (no subscript) and the spheroid (subscript e) terms.
In the above representation the singular point of the original kernel, located at $(s',\theta')=(s,\theta)$, cancels with the singular point of the spheroid kernel, located at $(s',\theta') = (s_e,\theta)$. 

The cancellation of the singular point in Eq.~\eqref{aSLBI} allows the integrands to be expanded. { This expansion should move all the $\theta'$ dependence to the numerators of the integrands so that the angular dependence can separated from the leading order term \cite{Koens2018}. This can be done with the binomial series,
\begin{equation}
(1+x)^\alpha = \sum_{k=0}^{\infty} \left(\begin{array}{c}
\alpha \\
 k
\end{array} \right) x^k,
\end{equation} 
where the generalised binomial coefficient is
\begin{equation}
\left(\begin{array}{c}
\alpha \\
 k
\end{array} \right)  =  \frac{1}{k!} \prod_{n=0}^{k+1} (\alpha-n). \label{binomial}
\end{equation}
This series converges absolutely for $|x|<1$ and $\alpha \in \mathbb{C}$ as the absolute value of each term is smaller than the previous. The binomial series can be applied to the denominators of our integrands by defining $|\R_{(e)}|^2 = |\tR_{(e)}|^2\left(1+\delta\R_{(e)}\right)$ where
\begin{eqnarray}
|\tR_{(e)}|^2&=&\R_{0(e)}^2 + \epsilon^2 \rho_{(e)}(s_{(e)})^2 +\epsilon^2 \rho_{(e)}(s')^2 \label{mod} \\
|\tR_{(e)}|^2 \delta\R_{(e)} &=& - 2\epsilon^2 \rho_{(e)}(s_{(e)}) \rho_{(e)}(s')\erho(s_{(e)},\theta) \cdot \erho(s',\theta') \notag \\
&& +2 \epsilon \rho_{(e)}(s_{(e)})  \R_{0(e)}\cdot \erho(s_{(e)},\theta) \notag \\
&& - 2 \epsilon \rho_{(e)}(s') \R_{0(e)}\cdot \erho(s',\theta') \label{cross}
\end{eqnarray}
Eq.~\eqref{mod} is the sum of the squared lengths of each vector within $\R_{(e)}$ and does not depend on $\theta'$ while Eq.~\eqref{cross} is the cross-vector terms occurring within $|\R_{(e)}|^2$ and so has dependence on $(s,s',\theta,\theta')$. The triangle inequality tells us that $|\R_{(e)}|^2 \leq  |\tR_{(e)}|^2$ and so $|\delta\R_{(e)}| \leq 1$. Furthermore $|\delta\R_{(e)}| = 1$ iff $(s',\theta') = (s,\theta)$, but this point was removed by the above regularisation. Hence we can produce a binomial series in $\delta\R_{(e)}$ to express} the boundary integral as
\begin{eqnarray}
8 \pi \mu \mathbf{U}(s,\theta)&=& \sum_{k=0}^{\infty} \int_{-1}^{1} \,ds' \int_{-\pi}^{\pi} \,d\theta' \delta\R^{k} \mathbf{G}^{(k)} \cdot \f(s',\theta') \notag \\
&& -\sum_{k=0}^{\infty} \int_{-1}^{1} \,ds' \int_{-\pi}^{\pi} \,d\theta' \delta\R_{e}^{k} \mathbf{G}^{(k)}_{e} \cdot \f(s,\theta)  \notag \\
&& + { \mathbf{M}_{A}'(s,\theta)}\cdot \f(s,\theta),  \label{expand}
\end{eqnarray}
where  $\R_{0(e)}=\rc_{(e)}(s_{(e)})-\rc(s')$,
\begin{eqnarray}
\mathbf{G}^{(k)}_{(e)}{(s',\theta',s,\theta)} &=&\left(\begin{array}{c}
- \frac{1}{2}\\
 k
\end{array} \right)\frac{\mathbf{I}}{|\tR_{(e)}|}  + \left(\begin{array}{c}
- \frac{3}{2}\\
 k
\end{array} \right)\frac{\R_{(e)} \R_{(e)}}{|\tR_{(e)}|^3}. \notag \\
\end{eqnarray}
 This binomial series converges absolutely provided $\delta\R_{(e)}<1$. This is true everywhere except at the singular point $(s',\theta')=(s,\theta)$. At this point, however, the individual terms from the body and spheroid cancel. Hence this expanded representation is exact.

{ The $k=0$ terms in the above series are larger than any other term in the summation. This is because, when $|\delta\R_{(e)}|<1$, each subsequent term in the series is smaller than the last, and when $|\delta\R_{(e)}|=1$ each $k$ of the body cancels with its spheroid counterpart.} Hence we { could} assume that the $k\neq 0$ and the angular dependent terms within $k=0$ are `relatively small' { and group them together. This turn the equation into
\begin{eqnarray}
8 \pi \mu \mathbf{U}(s,\theta)&=& \int_{-1}^{1} \,ds' \int_{-\pi}^{\pi} \,d\theta' \left[\mathbf{K} \cdot \f(s',\theta')-  \mathbf{K}_{e} \cdot \f(s,\theta)\right] \notag \\
&& + \varepsilon \Delta\mathcal{L}[\f]{(s,\theta)} +{ \mathbf{M}_{A}'(s,\theta)}\cdot \f(s,\theta), 
\end{eqnarray}
where
\begin{eqnarray}
 \mathbf{K}_{(e)} &=& \frac{\mathbf{I}}{|\tR_{(e)}|} + \frac{\R_{0(e)} \R_{0(e)}}{|\tR_{(e)}|^3},\\
\varepsilon \Delta\mathcal{L}[\f]{(s,\theta)} &=& \int_{-1}^{1} \,ds'  \int_{-\pi}^{\pi} \,d\theta' \left(\mathbf{G} - \mathbf{K} \right) \cdot \f(s',\theta') \notag\\
&& -\int_{-1}^{1} \,ds' \int_{-\pi}^{\pi} \,d\theta' \left( \mathbf{G}_{e}-\mathbf{K}_{e} \right) \cdot \f(s,\theta),  \notag \\
\end{eqnarray}
and $\varepsilon$ is the `small' parameter representing the angular and $k\neq 0 $ terms. In the above, $\varepsilon \Delta\mathcal{L}[\f]$ is a surface integral operator that contains all the $k\neq0$ and angular terms that are assumed `small'. It is equivalent to the boundary integral of the tubular body without the corresponding $k=0$ term minus the boundary integral of the spheroid without the $k=0$ term because of the convergence of the binomial series. The latter interpretation is the easier way to evaluate this term when $\f$ is known. It is not clear, a priori, that $\varepsilon \Delta\mathcal{L}[\f]$ is small but we note that its integrand cancels as $(s',\theta')\to (s,\theta)$ and is small when $(s',\theta')$ far from $(s,\theta)$ ($\delta R \ll 1$). This representation suggests
expanding} $\f(s,\theta)$ as
\begin{equation}
\f(s,\theta) = \sum_{n=0}^{\infty} (-1)^n {\varepsilon^n} \f_n(s,\theta), \label{expand}
\end{equation}
{and collecting like powers of $\varepsilon$. Hence the $\f_n(s,\theta)$ satisfy}
\begin{eqnarray}
8 \pi \mu \mathbf{U}(s,\theta)   &=& \int_{-1}^{1} \,ds' \mathbf{K} {(s',s)}\cdot \left\langle\f_0(s',\theta')\right\rangle_{\theta'}  \notag \\ 
&& +\mathbf{M}_{A}{(s,\theta)}\cdot \f_0(s,\theta), \label{f0}   \\
\Delta\mathcal{L}[\f_{n-1}]{(s,\theta)} &=&   \int_{-1}^{1} \,ds' \mathbf{K} {(s',s)} \cdot \left\langle\f_n(s',\theta')\right\rangle_{\theta'} \notag \\ 
&& + \mathbf{M}_{A}{(s,\theta)}\cdot \f_n(s,\theta). \label{f1}
\end{eqnarray}
Equations~\eqref{expand}, \eqref{f0} and \eqref{f1} { form} the tubular-body theory equations.
In them we have evaluated the { $k=0$ spheroid integrals} \cite{Gradshteyn2000} and defined $\left\langle(\cdot) \right\rangle_{\theta'} = \int_{-\pi}^{\pi} (\cdot) \,d\theta'$,
\begin{eqnarray}
\mathbf{M}_{A}{(s,\theta)} &=& \left\{\zeta_{\parallel}{(s,\theta)} \Th(s) \Th(s) +\zeta_{\perp}{(s,\theta)} [\mathbf{I}-\Th(s)\Th(s)]\right\}, \notag \\ \\
\zeta_{\parallel}{(s,\theta)} &=& \zeta_{\parallel}'  - \frac{1-\beta}{a (-\beta)^{3/2}}L-g{(1,s,\theta)}+g{(-1,s,\theta)}, \notag \\ \\
\zeta_{\perp}{(s,\theta)} &=&  \zeta_{\perp}'- \frac{1}{a \sqrt{-\beta}}L \\
L{(s,\theta)} &=& \ln\left(\frac{a(s_e -\beta) + \sqrt{-\beta} |\tR_e|(-1)}{a(s_e +\beta) + \sqrt{-\beta} |\tR_e|(1)} \right), \\
g{(s',s,\theta)} &=& \frac{2 (s_e-s')}{\beta |\tR_e|(s',s,\theta)}\left(\frac{  s' s_e \alpha^2 - (1-s_e^2)\beta}{2\beta - s_e^2 (1-\beta) } \right),
\end{eqnarray}
 Importantly, this series representation is exact and converges provided $|\f_{n-1}| > |\f_n|$. The proof of this is beyond the scope of this letter but we note that this condition was satisfied in all our tests as $\varepsilon \Delta\mathcal{L}[\f](s,\theta)$ was found to be small.

The TBT integral operator, present in Eqs.~\eqref{f0} and \eqref{f1}, is similar to a SBT operator but involves both $s$ and $\theta$. However averaging over this $\theta$ component it can be rewritten as
\begin{eqnarray}
 \left\langle \mathbf{M}_{A}^{-1}\cdot \mathbf{q}(s,\theta)\right\rangle_{\theta}  &=&\left\langle \mathbf{M}_{A}^{-1}\right\rangle_{\theta} \cdot\int_{-1}^{1} \,ds' \mathbf{K} \cdot \left\langle\f(s',\theta')\right\rangle_{\theta'} \notag \\
 && + \left\langle \f(s,\theta')\right\rangle_{\theta'}, \label{FH1}\\
\f(s,\theta) &=& \mathbf{M}_{A}^{-1} \cdot  \mathbf{q}(s,\theta) \notag \\
&& -\mathbf{M}_{A}^{-1} \cdot \left\langle \mathbf{M}_{A}^{-1}\right\rangle_{\theta}^{-1} \cdot   \left\langle \mathbf{M}_{A}^{-1}\cdot \mathbf{q}(s,\theta)\right\rangle_{\theta}  \notag \\
&& +\mathbf{M}_{A}^{-1} \cdot \left\langle \mathbf{M}_{A}^{-1}\right\rangle_{\theta}^{-1} \cdot  \left\langle \f(s,\theta')\right\rangle_{\theta'} \label{FH2}
\end{eqnarray}
where $\mathbf{q}(s,\theta)$ { represents the known functions on the left-hand side of Eqs.~\eqref{f0} and \eqref{f1} }. The above expression shows that solutions to the TBT operator is equivalent to a Fredholm integral equation of the second kind and a sequence of matrix operations. Furthermore the kernel of this integral operator is compact and self-adjoint. These kinds of Fredholm equations are well posed and diagonalisable \cite{Polianin2008}. As such, there are several ways to solve these problems numerically and analytical solutions can be expressed as in infinite series of integrals (see the supplementary material (SM) \cite{SM1,Polianin2008}). The structure of these kernels is also close to the modified SBT of Tornberg and Shelley \cite{Tornberg2004} and the SBT of Andersson \textit{et al.} \cite{Andersson2020} both of which have positive definite eigenvalues. 

\textit{Validation}- We solved the TBT integral equations, Eqs.~\eqref{f0} and \eqref{f1}, with a collocation method \cite{Polianin2008} and truncate the series in Eq.~\eqref{expand} at $n=N$. Briefly the collocation method divides $s\in[-1,1]$ into segments and assumes $\f$ is constant over each segment. Equation~\eqref{FH1} is then turned into a system of linear equations by enforcing that they hold at the centre of each segment (full details
provided in SM\cite{SM2, Polianin2008, Kim2005, MATLAB:2021}).  We note that this simple collocation implementation is possible because the integral kernel in Eq.~\eqref{FH1} is compact unlike in Keller and Rubinow's SBT.

\begin{figure}
\centering
\includegraphics[width=0.45\textwidth]{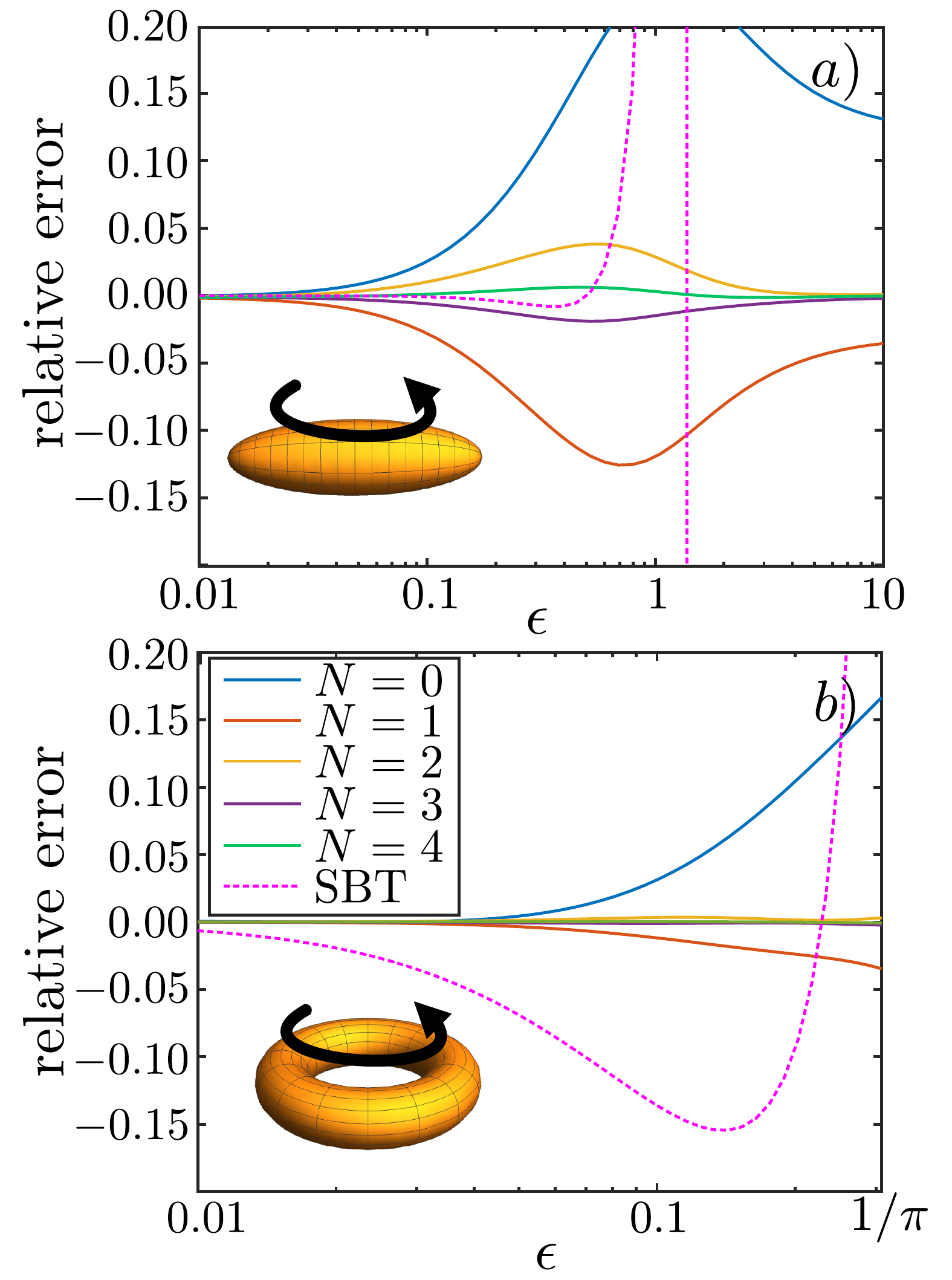}
\caption{The relative error between TBT and exact solution for torque from the broadwise rotation of a prolate spheroid (a) and the axisymmetric rotation or torus (b). The relative error is defined as the difference between the prediction and the exact coefficient divided by the exact coefficient. Different lines correspond to the TBT prediction with different levels of truncation; blue is the leading term ($N=0$), red is the first two terms ($N=1$), yellow is the first three ($N=2$), purple is the first four ($N=3$), and green is the first five ($N=4$). The pink dashed line is the SBT approximation.} 
\label{fig:validation}
\end{figure} 

The accuracy of the TBT equations, Eqs~\eqref{expand}, \eqref{f0} and \eqref{f1}, was tested against the solutions for the drag on a spheroid \cite{LAMB1932} and torus \cite{Pell2010,Kanwal2006, Majumdar1977, Pell2010} for a wide range of $\epsilon$. The terms with the largest relative error are shown in Fig.~\ref{fig:validation} (other terms available in SM\cite{SM3,LAMB1932,Koens2014,Kanwal2006, Majumdar1977, Pell2010,Johnson1979a,DORREPAAL1976, Majumdar1979}). The translation coefficients for the spheroid was found to capture the behaviour exactly over the entire range considered. This is because we used the solution for a translating spheroid as the effective geometry. The relative error between the known solutions and that predicted by TBT decreases as the number of terms kept in Eq.~\eqref{expand} increases (Fig.~\ref{fig:validation}). At leading order ($N=0$) the largest error is around 20\% and occurs when the spheroid is almost spherical, $\epsilon\sim 1$ and the torus is closed $\epsilon = 1/\pi$. Both of these cases lie well outside the typical SBT limits. When $\epsilon\sim 1$ the body is not slender while when $\epsilon = 1/\pi$ the curvature of the torus equals the thickness. This relative error decreases to less than 1\% after including the first five terms ($N=4$) in Eq.~\eqref{expand} for the entire region tested. The convergence of TBT, over a range much larger then available to classical SBT, suggests that TBT could prove effective when determining the viscous hydrodynamics of cable-like bodies in complex situations.

The power of TBT can be further demonstrated using a tightly wound helix (Fig.~\ref{fig:ex}). Helices are iconic in slow-viscous flow and occur in many mechanical and biological situations. Though common, relatively little is known about how the dynamics of these helices change as they become tightly wound, even though some micro-organisms form these tightly wound helical shapes \cite{Young2006}. One possible explanation of this is because such shapes lie well outside the SBT limits and so full numerical simulations would be needed. TBT, however, is exact and so can be used in these limits.

\begin{figure}
\centering
\includegraphics[width=0.45\textwidth]{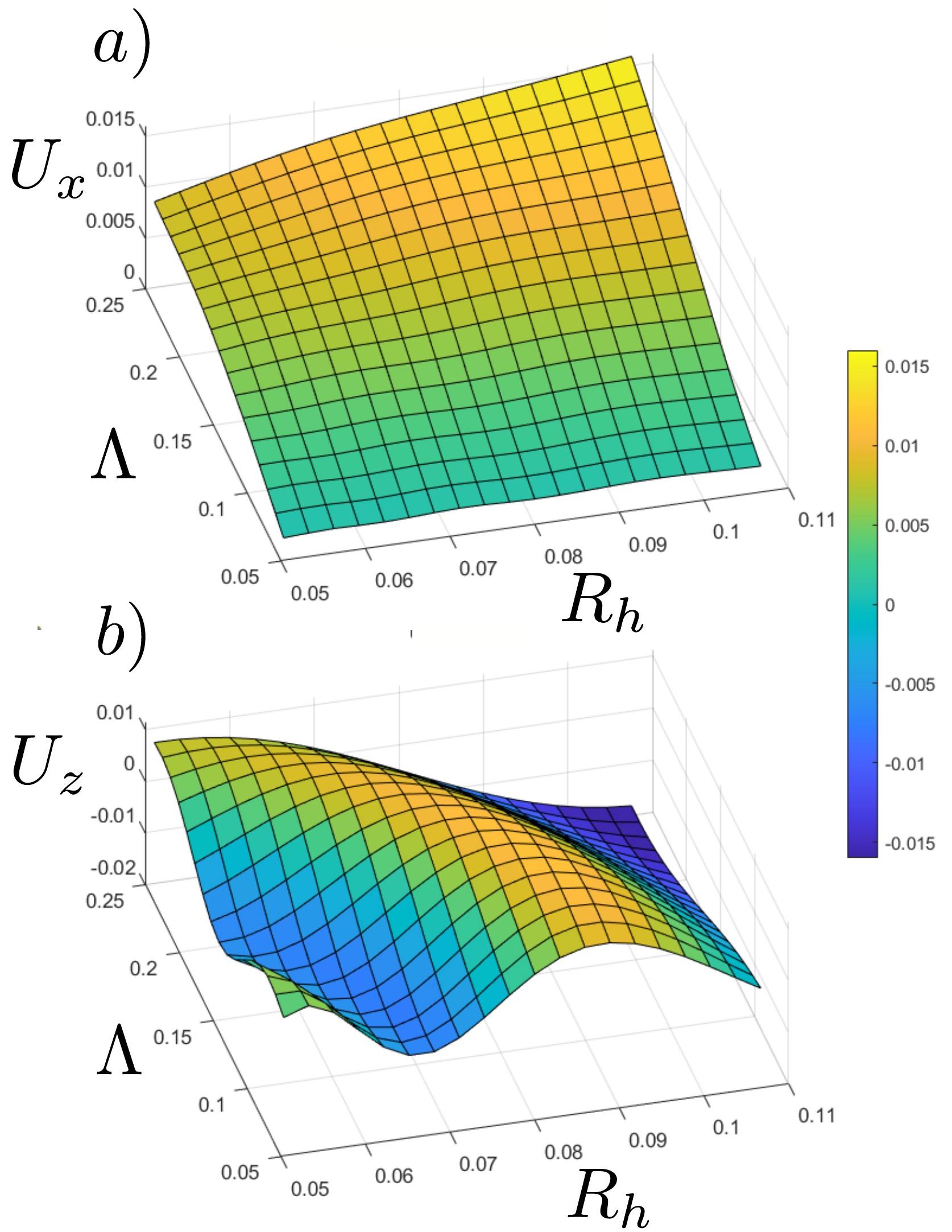}
\caption{The translational velocity of a force free helix from rotation around its axis ($x$-axis). a) velocity along its axis ($U_x$),  b) velocity perpendicular to axis ($U_z$). All plots are shown for $\epsilon=0.05$.} 
\label{fig:helix}
\end{figure} 

We investigated the velocity of a force free helix which is rotated around its central axis (Fig.~\ref{fig:helix}), varying the helix pitch, $\Lambda$ and radius, $R_h$ (Fig.~\ref{fig:ex}). The parametrisation used and resistance matrix of said helix is available in the SM\cite{SM4}. In this example we restricted ourselves to inextensible helices with $\epsilon=0.05$. The geometric constraint that the helix does not touch itself means $\Lambda, R_h >\epsilon$. For these bodies, Eq.~\eqref{expand} had converged by $N=6$. Over the range tested, the velocity of the helix along its axis increases as $\Lambda$ and  $R_h$ increases (Fig.~\ref{fig:helix}a). This is probably due to the coupling between translation and rotation increasing as the helix becomes less coiled. The velocity perpendicular to the helix axis, however, displays non-trivial oscillatory behaviour (Fig.~\ref{fig:helix}b). Hence, though these bodies are very tightly wound, the system still experiences non-trivial effects from its helical shape.

\textit{Discussion and Conclusion}- Fibre-like objects occur in many interesting and important situations but can be tricky to model theoretically. Direct numerical simulations typically require high levels of resolution, while the asymptotic slender-body theories are restricted to leading order solutions in specific regimes. The extension of these SBTs beyond these approximation limits has therefore been identified as a key problem for several interdisciplinary research fields \cite{Lauga2016, Gaffney2011, Nazockdast2017a, Ruzicka2011, Sanjay2018, DuRoure2019, Brennen1977a, Kugler2020, Tornberg2004, Reis2018}. 

This letter develops tubular-body theory; a SBT-like theory that can determine the traction on cable-like bodies exactly. TBT expresses the surface traction on the body as the sum of solutions to a one dimensional integral equation (Eqs.~\eqref{expand}, \eqref{f0} and \eqref{f1}). This integral equation is similar to that developed in SBT except with a compact and self-adjoint integral operator. These types of integral equations are called Fredholm integral equations of the second kind and have been studied extensively. Furthermore Fredholm integral equations of the second with compact and self-adjoint operators are well-posed and diagonalisable. As such there exists several methods to solve them both numerically and theoretically \cite{Polianin2008}.

We used a collocation method to numerically solve these equations. The compact nature of the integral operator means this representation does not suffer from issues with the kernel blowing up. We then compared the predictions of TBT to the exact solutions for drag on a spheroid and a torus. In both cases we found that, with only the first five terms in the series, TBT was able to determine the resistance coefficients to within 1\% error. This was tested for spheroids ranging from the very slender/prolate ($\epsilon=0.01$) to very oblate ($\epsilon=10$) and tori ranging from slender ($\epsilon=0.01$) to closed ($\epsilon=1/\pi$). We then used TBT to look at the velocity of a rotating force-free tightly-wound helix, observing that even in these tightly wound shapes the helix showed non-trivial dependence on the geometry.

In addition to being exact, TBT has several useful properties and its derivation can be generalised to more complicated geometries. The ability of TBT to resolve the traction on the body means the force and torque on the body in any flow can be determined from the Lorentz reciprocal relationship and the results for rigid body motion \cite{Kim2005}. This process is useful for determining the swimming of microscopic organisms and asymptotically interacting bodies through the method of reflections \cite{Kim2005, Brenner1963}. Furthermore the equations can be extended to filamentous bodies near (but not touching) walls or other objects by including the appropriate corrections to the single-layer boundary integral representation \cite{Pozrikidis1992}. This is because these terms do not introduce any new singular points in the Green's function when evaluated on the surface of the body and so the binomial series treatment holds. An improved effective geometry, Eq.~\eqref{effective}, may however improve the convergence of the final series representation.

Finally we note that, though we have focused on fibres in viscous flow, the derivation can be easily repeated for several other systems. The derivation itself relies on three elements: a boundary integral representation for the system, the known solution for spheroidal geometries and the binomial series. Thankfully many systems of interest, like diffusion, satisfy these conditions, with the behaviour of ellipsoids often being a solved geometry.

\acknowledgments{ LK was funded by Australian Research Council (ARC) under the Discovery Early Career Research Award scheme (grant agreement DE200100168).}

\bibliographystyle{ieeetr}
\bibliography{references}

\begin{thebibliography}{10}

\bibitem{Lauga2016}
E.~Lauga, ``{Bacterial Hydrodynamics},'' {\em Annu. Rev. Fluid Mech.}, vol.~48,
  pp.~105--130, 2016.

\bibitem{Gaffney2011}
E.~A. Gaffney, H.~Gad{\^{e}}lha, D.~J. Smith, J.~R. Blake, and J.~C.
  Kirkman-Brown, ``{Mammalian Sperm Motility: Observation and Theory},'' {\em
  Annu. Rev. Fluid Mech.}, vol.~43, pp.~501--528, 2011.

\bibitem{Nazockdast2017a}
E.~Nazockdast, A.~Rahimian, D.~Zorin, and M.~Shelley, ``{A fast platform for
  simulating semi-flexible fiber suspensions applied to cell mechanics},'' {\em
  J. Comput. Phys.}, vol.~329, pp.~173--209, 2017.

\bibitem{Ruzicka2011}
B.~Ruzicka and E.~Zaccarelli, ``{A fresh look at the Laponite phase diagram},''
  {\em Soft Matter}, vol.~7, p.~1268, 2011.

\bibitem{Sanjay2018}
M.~R. Sanjay, P.~Madhu, M.~Jawaid, P.~Senthamaraikannan, S.~Senthil, and
  S.~Pradeep, ``{Characterization and properties of natural fiber polymer
  composites: A comprehensive review},'' {\em J. Clean. Prod.}, vol.~172,
  pp.~566--581, 2018.

\bibitem{Lemma2021}
L.~M. Lemma, M.~M. Norton, A.~M. Tayar, S.~J. DeCamp, S.~A. Aghvami, S.~Fraden,
  M.~F. Hagan, and Z.~Dogic, ``{Multiscale Microtubule Dynamics in Active
  Nematics},'' {\em Phys. Rev. Lett.}, vol.~127, p.~148001, 2021.

\bibitem{Koens2021a}
L.~Koens and T.~D. Montenegro-Johnson, ``{Local drag of a slender rod parallel
  to a plane wall in a viscous fluid},'' {\em Phys. Rev. Fluids}, vol.~6,
  p.~064101, 2021.

\bibitem{Kim2005}
S.~Kim and S.~J. Karrila, {\em {Microhydrodynamics: Principles and Selected
  Applications}}.
\newblock Boston: Courier Corporation, 2005.

\bibitem{DuRoure2019}
O.~du~Roure, A.~Lindner, E.~N. Nazockdast, and M.~J. Shelley, ``{Dynamics of
  flexible fibers in viscous flows and fluids},'' {\em Annu. Rev. Fluid Mech.},
  vol.~51, pp.~539--572, 2019.

\bibitem{Tornberg2004}
A.~Tornberg and M.~J. Shelley, ``{Simulating the dynamics and interactions of
  flexible fibers in Stokes flows},'' {\em J. Comput. Phys.}, vol.~196,
  pp.~8--40, 2004.

\bibitem{Keller1976a}
J.~B. Keller and S.~I. Rubinow, ``{Slender-body theory for slow viscous
  flow},'' {\em J. Fluid Mech.}, vol.~75, pp.~705--714, 1976.

\bibitem{Johnson1979}
R.~E. Johnson, ``{An improved slender-body theory for Stokes flow},'' {\em J.
  Fluid Mech.}, vol.~99, pp.~411--431, 1979.

\bibitem{Koens2018}
L.~Koens and E.~Lauga, ``{The boundary integral formulation of Stokes flows
  includes slender-body theory},'' {\em J. Fluid Mech.}, vol.~850, p.~R1, 2018.

\bibitem{Gotz2000}
T.~G{\"{o}}tz, {\em {Interactions of Fibers and Flow: Asymptotics, Theory and
  Numerics}}.
\newblock PhD thesis, University of Kaiserslautern, Kaiserslautern, Germany,
  2000.

\bibitem{Andersson2020}
H.~I. Andersson, E.~Celledoni, L.~Ohm, B.~Owren, and B.~K. Tapley, ``{An
  integral model based on slender body theory, with applications to curved
  rigid fibers},'' {\em Phys. Fluids}, vol.~33, p.~041904, 2021.

\bibitem{Barta1988}
E.~Barta and N.~Liron, ``{Slender Body Interactions for Low Reynolds
  Numbers—Part I: Body-Wall Interactions},'' {\em SIAM J. Appl. Math.},
  vol.~48, pp.~992--1008, 1988.

\bibitem{Gueron1992}
S.~Gueron and N.~Liron, ``{Ciliary motion modeling, and dynamic multicilia
  interactions},'' {\em Biophys. J.}, vol.~63, pp.~1045--1058, 1992.

\bibitem{Higdon2006}
J.~J.~L. Higdon, ``{A hydrodynamic analysis of flagellar propulsion},'' {\em J.
  Fluid Mech.}, vol.~90, p.~685, 1979.

\bibitem{Man2016}
Y.~Man, L.~Koens, and E.~Lauga, ``{Hydrodynamic interactions between nearby
  slender filaments},'' {\em EPL (Europhysics Lett.}, vol.~116, p.~24002, 2016.

\bibitem{Brennen1977a}
C.~Brennen and H.~Winet, ``{Fluid Mechanics of Propulsion by Cilia and
  Flagella},'' {\em Annu. Rev. Fluid Mech.}, vol.~9, pp.~339--398, 1977.

\bibitem{Katsamba2020}
P.~Katsamba, S.~Michelin, and T.~D. Montenegro-Johnson, ``{Slender Phoretic
  Theory of chemically active filaments},'' {\em J. Fluid Mech.}, vol.~898,
  p.~A24, 2020.

\bibitem{Kugler2020}
S.~K. Kugler, A.~Kech, C.~Cruz, and T.~Osswald, ``{Fiber Orientation
  Predictions—A Review of Existing Models},'' {\em J. Compos. Sci.}, vol.~4,
  p.~69, 2020.

\bibitem{Reis2018}
P.~M. Reis, F.~Brau, and P.~Damman, ``{The mechanics of slender structures},''
  {\em Nat. Phys.}, vol.~14, pp.~1150--1151, 2018.

\bibitem{Pozrikidis1992}
C.~Pozrikidis, {\em {Boundary Integral and Singularity Methods for Linearized
  Viscous Flow}}.
\newblock Cambridge University Press, 1992.

\bibitem{Batchelor2006}
G.~K. Batchelor, ``{Slender-body theory for particles of arbitrary
  cross-section in Stokes flow},'' {\em J. Fluid Mech.}, vol.~44, pp.~419--440,
  1970.

\bibitem{Brenner1963}
H.~Brenner, ``{The Stokes resistance of an arbitrary particle},'' {\em Chem.
  Eng. Sci.}, vol.~18, pp.~1--25, 1963.

\bibitem{Martin2019}
C.~P. Martin, ``{Surface tractions on an ellipsoid in Stokes flow: Quadratic
  ambient fields},'' {\em Phys. Fluids}, vol.~31, p.~021209, 2019.

\bibitem{Gradshteyn2000}
I.~S. Gradshteyn, I.~M. Ryzhik, A.~Jeffrey, and D.~Zwillinger, {\em {Table of
  Integrals, Series, and Products}}.
\newblock San Diego, California: Academic Press, 2000.

\bibitem{Polianin2008}
A.~D. Polianin and A.~V. Manzhirov, {\em {Handbook of integral equations}}.
\newblock Boca Raton, Florida: Chapman {\&} Hall/CRC, 2nd~ed., 2008.

\bibitem{SM1}
See Sec.~1 of the Supplemental Material at [URL will be inserted by publisher]
  for the explicit analytical solution to the tubular body equations.

\bibitem{SM2}
See Sec.~2 of the Supplemental Material at [URL will be inserted by publisher]
  for a detailed description of the numerical method.

\bibitem{MATLAB:2021}
MATLAB, {\em version 9.10.0.1710957 (R2021a)}.
\newblock Natick, Massachusetts: The MathWorks Inc., 2021.

\bibitem{LAMB1932}
H.~Lamb, {\em {Hydrodynamics}}.
\newblock Cambridge: Cambridge University Press, 6th~ed., 1932.

\bibitem{Pell2010}
W.~H. Pell and L.~E. Payne, ``{On Stokes flow about a torus},'' {\em
  Mathematika}, vol.~7, pp.~78--92, 1960.

\bibitem{Kanwal2006}
R.~P. Kanwal, ``{Slow steady rotation of axially symmetric bodies in a viscous
  fluid},'' {\em J. Fluid Mech.}, vol.~10, p.~17, 1961.

\bibitem{Majumdar1977}
S.~R. Majumdar and M.~E. O'Neill, ``{On axisymmetric stokes flow past a
  torus},'' {\em Zeitschrift f{\"{u}}r Angew. Math. und Phys. ZAMP}, vol.~28,
  pp.~541--550, 1977.

\bibitem{SM3}
See Secs.~4 and 5 of theSupplemental Material at [URL will be inserted by
  publisher] for the other resisatance coefficents for the sphereoid and the
  torus.

\bibitem{Koens2014}
L.~Koens and E.~Lauga, ``{The passive diffusion of Leptospira interrogans},''
  {\em Phys. Biol.}, vol.~11, p.~066008, 2014.

\bibitem{Johnson1979a}
R.~E. Johnson and T.~Y. Wu, ``{Hydromechanics of low-Reynolds-number flow. Part
  5. Motion of a slender torus},'' {\em J. Fluid Mech.}, vol.~95, pp.~263--277,
  1979.

\bibitem{DORREPAAL1976}
J.~M. Dorrepaal, S.~R. Majumdar, M.~E. O'Nejll, and K.~B. Ranger, ``{A Closed
  Torus in Stokes Flow},'' {\em Q. J. Mech. Appl. Math.}, vol.~29,
  pp.~381--397, 1976.

\bibitem{Majumdar1979}
S.~R. Majumdar and M.~E. O'Neill, ``{Asymmetric stokes flows generated by the
  motion of a closed torus},'' {\em Zeitschrift f{\"{u}}r Angew. Math. und
  Phys. ZAMP}, vol.~30, pp.~967--982, 1979.

\bibitem{Young2006}
K.~Young, ``{The selective value of bacterial shape.},'' {\em Microbiol. Mol.
  Biol. Rev.}, vol.~70, pp.~660--703, 2006.

\bibitem{SM4}
See Sec.~5 of the Supplemental Material at [URL will be inserted by publisher]
  for parametrisation used and resistance matrix of said helix is available.

\end{thebibliography}
\end{document}